\documentclass[traditabstract]{aa}
\usepackage{graphicx}
\usepackage{natbib}
\usepackage{txfonts}
\bibpunct{(}{)}{;}{a}{}{,}
\begin{document}

  \title{The main sequences of NGC2808: constraints on the early disc accretion scenario}
  \author{Santi Cassisi\inst{1} \and   
          Maurizio Salaris\inst{2}}

\institute{INAF~$-$~Osservatorio Astronomico di Collurania, Via M. Maggini, I$-$64100 , Teramo, Italy 
            \email{cassisi@oa-teramo.inaf.it} 
           \and Astrophysics Research Institute, 
           Liverpool John Moores University, 
           IC2, Liverpool Science Park, 
           146 Brownlow Hill, 
           Liverpool L3 5RF, UK
           }
  \abstract{A new scenario --early disc accretion-- has been proposed very recently  
to explain the origin of the multiple population phenomenon in Galactic globular clusters. It envisages 
the possibility that a fraction of low- and very low-mass cluster stars may accrete 
the ejecta of interacting massive binary (and possibly also fast rotating massive) 
stars during the fully convective, pre-main sequence stage, to reproduce the CN and ONa anticorrelations observed among stars in individual clusters.
This scenario is assumed to be able to explain the presence (and properties) of the multiple populations 
in the majority of globular clusters in the Milky Way. 
Here we have considered the well studied cluster NGC~2808, which displays a triple main sequence with well defined and separate 
He abundances. Knowledge of these abundances allowed us to put strong constraints 
on the He mass fraction and amount of matter to be accreted by low-mass pre-main sequence stars.   
We find that the minimum He mass fraction in the accreted gas has to be $\sim0.44$ to produce the observed 
sequences and that at fixed initial mass of the accreting star, different efficiencies for the accretion are required to produce stars placed 
onto the multiple main sequences. This may be explained by 
differences in the orbital properties of the progenitors and/or different spatial 
distribution of intracluster gas with varying He abundances. 
Both O-Na and C-N anticorrelations appear naturally along the main sequences, once considering 
the predicted relationship between He and CNONa abundances in the ejecta of the polluters. 
As a consequence of the accretion, we predict no discontinuity between the abundance ranges covered by intermediate and blue main sequence stars, 
but we find a sizeable (several 0.1~dex) discontinuity of the N and Na abundances 
between objects on the intermediate and red main sequences.
There is in principle enough polluting gas with the right He abundances to explain the observed main sequences by early accretion
(hence no need to invoke a much higher cluster mass at birth, as required by scenarios 
envisaging multiple star formation bursts), however 
the matter ejected by the polluters must not be fully mixed before being accreted otherwise,   
according to the models, the average He abundance 
of the polluting gas is too low to produce the observed multiple main sequences.
}
\authorrunning{Cassisi S. \& Salaris M.}
\titlerunning{NGC2808 and the early accretion scenario}
  \maketitle


\section{Introduction}

During the last decade, our understanding of the physical mechanism(s) driving Galactic globular cluster (GGC) formation
has been seriously challenged by the discovery of multiple stellar populations in individual GGCs 
\citep[see e.g.][and references therein for a review on 
this subject]{gratton:12, piotto:12}. 
In a nutshell, a GGC hosts distinct subpopulations with different chemical patterns involving the light elements C, N, O, Na 
and sometimes Mg and Al, plus He; 
depending on the photometric filters employed, these subpopulations may 
lie on separate sequences in the observed colour-magnitude-diagrams  
\citep[CMDs -- see e.g.][for a detailed discussion]{sbordone:11, cassisi:13}.
Whilst the multiple stellar population phenomenon affects most of the GGCs observed so far, 
every cluster appears to be somewhat unique, in the sense
that each individual GGC shows its own realization of the multiple population phenomenon.

The mainstream scenarios proposed to explain the formation of subpopulations within individual GGCs 
\citep{decressin, dercole:08,dercole:11,conroy:11, valcarce:11} postulate   
that the chemical patterns observed within each cluster are produced by multiple star formation episodes during the early stages of 
its evolution. After a first generation (FG) 
of stars with the chemical abundance ratios typical of the halo field population is formed, successive generations (SG stars) originate in 
matter ejected by pre-existing FG stars ({\sl polluters}), diluted with material of FG composition not yet involved in star formation episodes.   
These scenarios differ in terms of the origin of the polluters, which are identified as intermediate-mass asymptotic giant branch stars 
(IM-AGB stars) or  
fast-rotating massive star (FRMS) or interacting massive binary star (IMBS) systems. Each of these objects is in principle able to 
eject in the intracluster medium matter with varied C, N, O, Na, §Mg, Al and He abundances, compared to the FG composition, chemically processed 
by high-temperature proton captures during H-burning. The second difference, related to the nature of the polluters, is the amount of dilution 
necessary to explain the 
abundances of SG stars. 

Dilution with material with FG composition is necessary to produce the Na-O anticorrelation observed in SG stars (or to 
allow for the presence of Li in SG objects, if polluters are massive stars) out of the polluting gas, but 
no convincing physical mechanism(s) has (have) been envisaged yet to enable a star-forming cluster to 
keep a significant amount of cold FG gas. Another major difficulty is related to the available amount of polluting matter. 
SG stars make up $\sim$30-70\% 
of the total current stellar content in a GGC \citep{carretta:09}, but only a relatively small fraction of FG stars 
is contained in the candidate FG polluters for a {\sl normal} initial mass function. This requires that the initial mass of the present GGCs 
had to be higher by a factor of 10 to 100 at birth, which is at odds with empirical constraints 
coming from observations of the globular cluster system in the Fornax dwarf galaxy \citep[see e.g.][and references therein]{larsen:12}.
Also, as investigated by \citet{bcd}, 
there has so far been no evidence of ongoing star formation within Galactic and extragalatic young massive clusters.

Due to these outstanding issues, it is worthwhile exploring alternative avenues  
for the GGC formation, which do not require fine tuning for individual objects, once intrinsic properties such as 
current total mass, gravitational potential, and position within the Galaxy, are accounted for.

Alternative scenarios \citep{dgc:83, thoul:02} 
envisage that a fraction of low-mass FG stars during the main sequence (MS) accrete matter ejected by the polluters, and 
modify their surface composition 
to match the observed range of CNONaMgAl abundances. Actually there would not be additional episodes of star formation, just FG stars 
with modified surface abundances.
This accretion scenario is particularly attractive because it may avoid the need of 
dilution with FG gas (the dilution happens on the FG stars, within their convective envelopes) and does not require 
the cluster to be much more massive at birth (only the composition of the surface convective regions of a fraction of FG stars 
need to be affected, hence less polluting matter is required). 
The theoretical analysis 
by \cite{newsham:07} has shown that --at least in the case of $\omega$~Cen-- this scenario can in principle explain the presence of distinct sequences 
in the CMD \citep{king:12}, but spectroscopic observations of GGCs cannot be reproduced. The problem is that observed abundance pattern (i.e., the O-Na 
anticorrelation) is observed in both MS turn off stars and red giants. Given the large difference of the mass of the surface convection zone 
\citep{sc:05}, if only convective envelopes of MS stars display the observed O-Na anticorrelation, 
this pattern is completely washed out when stars climb the red giant branch. 

The accretion scenario has been recently modified by \citet{bastian:13}, who  
propose that accretion does occur during the pre-MS, when stars are fully convective --a so called 
{\sl early disc accretion}. This means that  
the polluting matter is diluted within the whole star, hence no further dilution occurs during the red giant branch phase.  
The cluster would be gas-free after a single star formation episode 
with a standard initial mass function, with the more massive stars (i.e. both FRMS and IMBS objects) strongly 
centrally concentrated as a consequence 
of mass segregation. The matter ejected by these fast evolving 
massive stars would then be quickly accreted onto the circumstellar discs surrounding low-mass 
objects still on their pre-MS. Accretion is actually favoured by 
the presence of a circumstellar disc, which increases the actual cross-section for this process. 
Since the ejecta of massive stars would be preferentially 
concentrated in the cluster core, only the fraction of low-mass stars that crosses these regions during their orbits will 
show O-Na anticorrelations. The extent of the anticorrelation is then modulated by the time spent in the cluster core, 
and the efficiency of accretion.
As a {\sl natural} consequence of this scenario, pollution cannot involve stars more massive than $\sim 1.5{\rm M_{\odot}}$, 
because their pre-MS lifetime would be shorter than the timescale for pollution from FRMS and IMBS objects.

Very importantly, recent hydrodynamical simulation by \citet{rp} suggest that core-collapse SN explosions may not be very efficient at removing 
from the cluster the dense ejecta of IMBS and FRMS objects, an important prerequisite for the viability of this scenario.    

According to the authors, this model should be valid for all GGCs that do not show a significant iron spread (most of the GGCs). 
In this paper we put first strong quantitative constraints on the efficiency of accretion and the He mass fraction in the accreted material, 
necessary  to reproduce the observed MS of NGC~2808 within this early disc accretion scenario. 
Analyses of the triple MS in the optical CMD, and the UV CMD of its horizontal branch stars 
have disclosed the presence of three sequences with different and well determined He abundances 
\citep{piotto:07, dalessandro, milone:12b}. By making use of Monte Carlo (MC) simulations, we derived strong 
constraints on the He mass fraction and amount of matter accreted by low-mass pre-MS stars, 
to reproduce the observed He abundance distribution within the cluster. By considering 
models for the polluters, we also studied the resulting C-N and O-Na anticorrelations.  
Our simulations are presented in Section~2, followed by a discussion of the results and their impact on the early disc accretion scenario.

\section{Simulations}
\label{data}

The starting point of our investigation is the present He-abundance distribution along the triple MS observed in the GGC NGC~2808.
Accurate fits to the cluster CMD by \citet{milone:12b} with BaSTI \citep{basti1, basti2} isochrones 
have provided Y=0.248, 0.32 and 0.38 for the red (rMS), intermediate (mMS) and blue (bMS), respectively. 
The use of independent isochrones \citep{dotter} and the analysis of the UV CMD of horizontal branch stars \citep{dalessandro} 
confirm these values. The He abundances, especially of the bMS,   
are somewhat extreme amongst the observed GGCs, which usually display smaller He abundance ranges. 
However, NGC2808 is probably the cluster where the 
He abundances of its populations are more precisely determined, a crucial prerequisite for the type of analysis we carried out, 
and it is also one of the most challenging objects for any of the scenarios proposed to explain the multiple population phenomenon.
Assuming that the mass function (MF) of the three sequences can be modelled 
as $dn/dm \propto m^{-\alpha}$, the slope $\alpha$ is the same between approximately the MS turn off 
and a lower limit of $\sim$0.6 ${\rm M_{\odot}}$, $\sim$0.5 ${\rm M_{\odot}}$ and $\sim$0.45 ${\rm M_{\odot}}$ for the 
rMS, mMS and bMS, respectively. The corresponding 
values are $\alpha=-1.2 \pm 0.3$, $\alpha=-0.9 \pm 0.3$ and $\alpha=-0.9 \pm 0.4$, respectively.   
According to \citet{milone:12b} there is a marginal evidence that the MF for the rMS flattens below 
$\sim$0.6 ${\rm M_{\odot}}$, but this is at the limit of the mass range covered by their analysis.

Our goal was to determine what range of accreted masses and He-abundances is compatible with the observed 
triple MS of this cluster, and what is the resulting distribution of accreting seed masses. 
This results in strict constraints on the early disc accretion scenario applied to this cluster.

To this purpose, we performed a series of MC simulations of a large sample (20000 objects for each sequence) of  
synthetic stars belonging to the mMS and bMS. 
We started by drawing randomly a value for both the actual He-abundance and MS mass ${\rm M_{MS}}$, this latter according 
to a MF proportional to ${\rm m^{-0.9}}$, consistent with the estimates for both sequences; the precise value of the exponent is 
not critical to this analysis. For the bMS we assumed ${\rm Y_{bMS}}=0.38 \pm 0.01$ 
whilst for the mMS ${\rm Y_{mMS}}=0.32 \pm 0.01$, both with Gaussian distributions. 
We considered these small Gaussian spreads based on the indication that the components of the 
multiple MS in the CMD, once photometric errors are acounted for, appear to have an intrinsic width \citep[see discussion in][]{milone:12b}. 

We considered a mass range between 0.45 and 0.66 ${\rm M_{\odot}}$ for the bMS stars, and 0.45 and 0.72 ${\rm M_{\odot}}$ 
for the mMS ones. The upper limits are the turn off masses for an age of 12~Gyr (they change by $\sim$0.03 ${\rm M_{\odot}}$ 
if the age increases to 14~Gyr), and the lower limits correspond approximately to the lower mass limit 
of the MF determinations.

For each pair of MS mass and Y values, and an assumed He abundance in the accreted matter ${\rm Y_{accr}}$ -- 
the {\sl free parameter} of our simulations--  
we determined the corresponding amount of mass accreted ${\rm \Delta M_{accr}}$ in the assumption that the accreting objects are pre-MS  
fully convective stars, according to

\begin{equation}
{\rm 
\Delta M_{accr}=M_{MS} \ \frac{Y_{MS}-0.248}{Y_{accr}-0.248}
}
\label{eq}
\end{equation}

where ${\rm Y_{MS}}$ denotes the actual Y abundance on either the bMS or the mMS.
This simple equation already tells us that, obviously, stars on the two sequences with the same actual mass ${\rm M_{MS}}$ 
must have accreted with different efficiency for a given value of ${\rm Y_{accr}}$.

We have considered two scenarios. The simplest one assumes that ${\rm Y_{accr}}$ was the same for all stars 
belonging to both bMS and mMS, whilst in the second case we account for a distribution of ${\rm Y_{accr}}$ values.

\subsection{Constant   ${\rm Y_{accr}}$}

It is easy to realize that there must be a lower limit for ${\rm Y_{accr}}$, to be able to reproduce both bMS and mMS.
This is set by the minimum value for the accretor mass ${\rm M_i}$, 
The value of this minimum mass, in the absence of other constraints, is to some extent a free parameter. 
In the calculations that follow 
we have employed a minimum ${\rm M_i}$ 
equal to 0.1${\rm M_{\odot}}$, which is approximately the minimum stellar mass produced in a star formation event. 
It is of course possible also to consider in principle lower values, in the regime of brown dwarfs. In our final discussion 
we show how this would not alter substantially the conclusions of this analysis. 

 \begin{figure}
\centering
\includegraphics[scale=.4000]{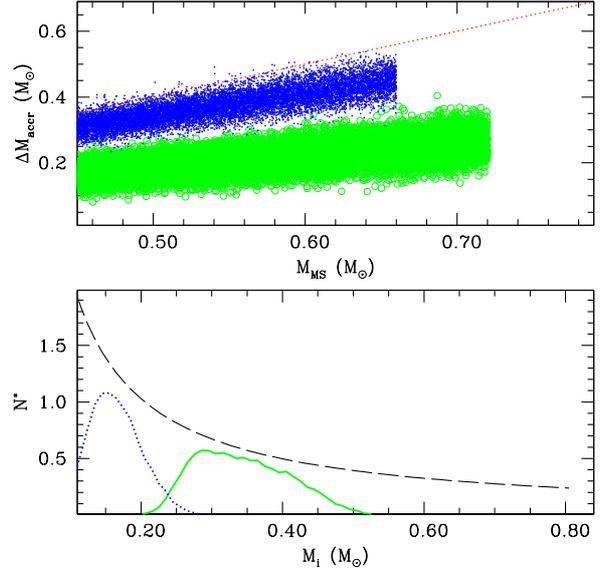}
\caption{{\sl Upper panel}: Mass accreted as a function of the actual mass (both in solar mass units)  
for mMS (green open circles) and bMS (blue dots) synthetic stars, and ${\rm Y_{accr}=0.44}$. The red dotted line 
is the upper bound of the region where the mass ${\rm M_i}$ of the accretors is $\ge$ 0.1${\rm M_{\odot}}$ (see text for details).
{\sl Lower panel}: 
Mass function of the accretors (bin size equal to 0.01 ${\rm M_{\odot}}$). 
The solid green line corresponds to the seeds of mMS stars, the dotted line is the counterpart for bMS stars. 
The quantity ${\rm N^{*}}$ corresponds to 
the number of objects with mass ${\rm M_i}$, normalized to the actual number of either bMS or mMS stars, 
multiplied by a factor 10. The dashed line displays a mass function $dn/dm \propto m^{-0.9}$ (see text for details).}
\label{sim1a}
\end{figure}

\begin{figure}
\centering
\includegraphics[scale=.4000]{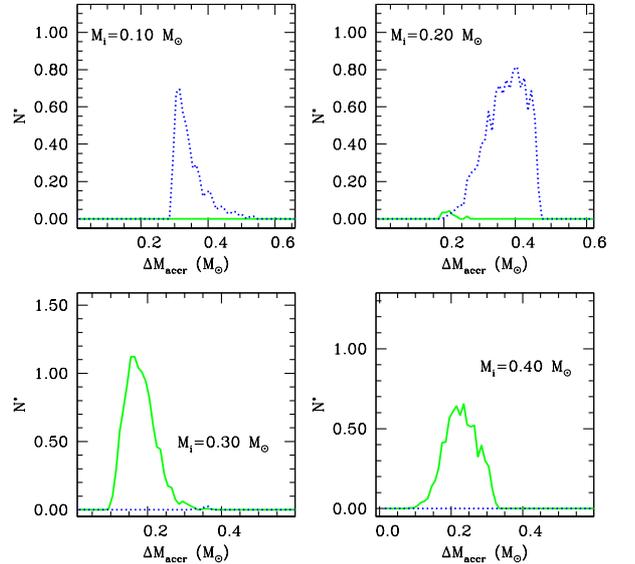}
\caption{Number distribution of accreted mass ${\rm \Delta M_{accr}}$ (bin size equal to 0.01 ${\rm M_{\odot}}$) for selected accretors with the 
labelled masses, and ${\rm Y_{accr}=0.44}$.
The quantity ${\rm N^{*}}$ corresponds to 
the number of objects that have accreted ${\rm \Delta M_{accr}}$ mass (solid green lines 
for mMS objects, dotted blue lines for bMS objects), normalized to the actual number of either bMS or mMS stars, 
multiplied by a factor 100.}
\label{sim1b}
\end{figure}

Figures~\ref{sim1a} and ~\ref{sim1b} show the results of our simulations 
for the case ${\rm Y_{accr}}$=0.44, which represents the sought lower limit. 
The top panel of Fig.~\ref{sim1a} displays ${\rm \Delta M_{accr}}$ as a function of ${\rm M_{MS}}$ for both mMS and bMS. 
At fixed ${\rm M_{MS}}$, there is a $\sim$0.1-0.2~${\rm M_{\odot}}$ range of ${\rm \Delta M_{accr}}$, due to the small spread 
of the present He abundance we assumed for both sequences. The different slopes in the ${\rm \Delta M_{accr}}$-${\rm M_{MS}}$ 
diagram are a consequence of Eq.~\ref{eq}. 
The dotted line corresponds to the relationship  ${\rm \Delta M_{accr}}$=${\rm M_{MS}}-$0.1, and 
points above this line correspond to MS stars originated from accretion on 
seeds with masses ${\rm M_i} <$0.1~${\rm M_{\odot}}$.
A physical interpretation for these MS stars in the simulation is that they are objects that cannot accrete enough matter to reach the 
observed He abundance, and will end up in between the three observed sequences.

Decreasing ${\rm Y_{accr}}$ below 0.44 
would move rigidly all points upwards in the diagram, because more matter needs to be accreted 
to attain the observed Y at fixed ${\rm M_{MS}}$. 
As a consequence, increasing numbers of bMS objects --and eventually mMS stars too-- will cross the 
boundary ${\rm M_i} =$0.1~${\rm M_{\odot}}$, starting from the lowest ${\rm M_{MS}}$, towards larger values. 
This implies that when ${\rm Y_{accr}}$ decreases below 0.44,  
the bMS would appear truncated at a mass (hence magnitude) where we still observe  
three well defined sequences. In the framework of the early disc accretion scenario, this occurrence sets a very strong constraint
on the minimum helium abundance of the accreted matter, provided by the \lq{polluters}\rq\, necessary 
to form the observed bMS.

The lower panel of the same Fig.~\ref{sim1a} displays the MF of the accretors for mMS and bMS stars, which 
in a real cluster must 
result from the combination of the initial MF of the accretors, plus the efficiency of the accretion process. 
It is noticeable the difference between the two MFs, which show a different shape, peak at different 
values of ${\rm M_i}$, and span different and essentially non-overlapping mass ranges. 
The peaks of the MF are at $\sim$0.15${\rm M_{\odot}}$ for the bMS, and 
$\sim$0.30${\rm M_{\odot}}$ for the mMS. This latter MF is overall flatter and spans a larger range of  ${\rm M_i}$.
For the sake of comparison we display also the functional form of the actual MF for both mMS and bMS. 
Overall, in the mass range of our simulations, $\sim$68\% of the actual bMS mass must have been accreted if ${\rm Y_{accr}}$=0.44, 
this fraction decreasing to $\sim$37\% for the mMS.

Figure~\ref{sim1b} displays the distribution of ${\rm \Delta M_{accr}}$ for selected values of ${\rm M_i}$. It is clear that 
the efficiency of accretion must be strongly dependent on ${\rm M_i}$.

\begin{figure}
\centering
\includegraphics[scale=.4000]{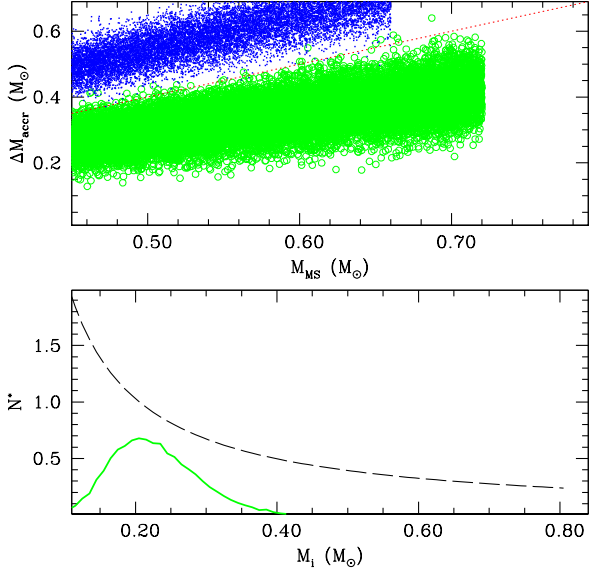}
\caption{As Fig.~\ref{sim1a} but for the case of ${\rm Y_{accr}}$=0.37.}
\label{sim1ab}
\end{figure}

Assuming --in a probably unrealistic case-- that mMS and bMS are produced by two completely separate accretion events, ${\rm Y_{accr}}$=0.37 is 
the minimum value necessary for producing 
the observed mMS (obviously the bMS cannot be formed with this {\sl low } value of ${\rm Y_{accr}}$), 
as displayed by Fig.~\ref{sim1ab}. For this case $\sim$58\% of the actual mMS mass must have been accreted.

\begin{figure}
\centering
\includegraphics[scale=.4000]{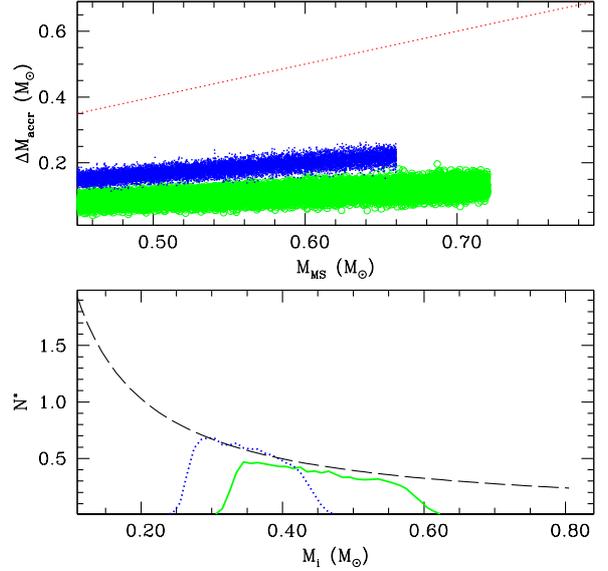}
\caption{As Fig.~\ref{sim1a} but for the case of ${\rm Y_{accr}}$=0.64.}
\label{sim1ac}
\end{figure}

We close this section with the case ${\rm Y_{accr}}$=0.64, which is the maximum He mass fraction in the ejecta of 
the IMBS model by \citet{demink}, the benchmark used by \citet{bastian:13}. Figure~\ref{sim1ac} shows how the mass 
of the accretors obviously increases compared to the previous cases, for both mMS and bMS. There is now a 
substantial overlap between 
the range of accretor masses for the two sequences, implying that some mechanism (maybe different orbital parameters) should 
{\sl tune} the accretion rate, given that the same ${\rm M_i}$ must obviously accrete 
different specific amounts of matter to populate either the mMS or the bMS.  
In the mass range of our simulations, $\sim$33\% of the actual bMS mass must have been accreted if ${\rm Y_{accr}}$=0.64, 
this fraction decreasing to $\sim$18\% for the mMS.

\subsection{Variable   ${\rm Y_{accr}}$}

Most probably the case of accretion of matter with constant ${\rm Y_{accr}}$ is not realistic, given that 
both IMBS and FRMS ejecta are expected to display a range of He values, unless they  
are well mixed within extremely short timescales, or --very improbably-- for some reason 
stars accrete selectively only matter with a single He abundance.
Obviously, if matter with ${\rm Y_{accr}}>$0.44 is accreted, according to the previous simulations it is always 
possible, with the appropriate accretion rates, to produce the observed mMS and bMS.  

\begin{figure}
\centering
\includegraphics[scale=.4000]{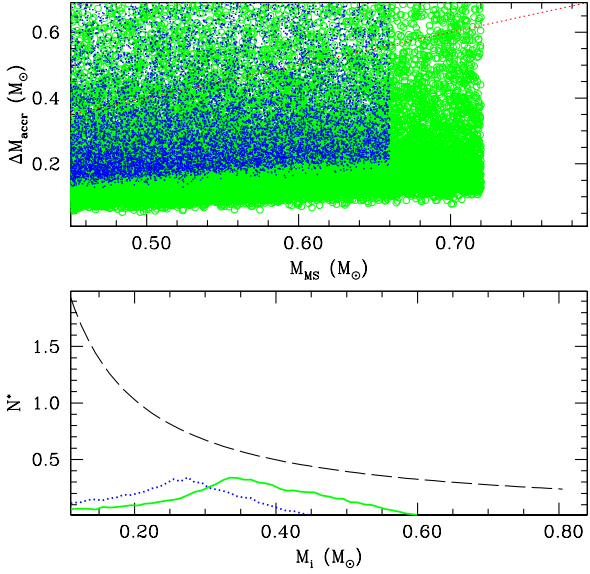}
\caption{As Fig.~\ref{sim1a} but for the case of uniform probability to accrete matter with ${\rm Y_{accr}}$ between 0.25 and 0.64 
(see text for details).}
\label{sim2a}
\end{figure}

We considered the full range of He abundances in the matter ejected by \citet{demink} IMBS models, e.g. 
Y between 0.25 and 0.64. A simple 
uniform distribution for the probability of a current bMS or mMS star to have accreted gas with  ${\rm Y_{accr}}$ 
within this range, cannot produce the observed bMS and mMS, as shown by Fig.~\ref{sim2a}.
Over 30\% of bMS synthetic stars and $\sim$15\% of mMS objects, spread along the whole 
simulated mass range, are unable to accrete enough matter. This would imply a large population of stars spread between the 
multiple CMD sequences, that would essentially be blurred into one.
The mean value of ${\rm Y_{accr}}$ in this case is 0.45, with a 1$\sigma$ dispersion equal to 0.11.

\begin{figure}
\centering
\includegraphics[scale=.4000]{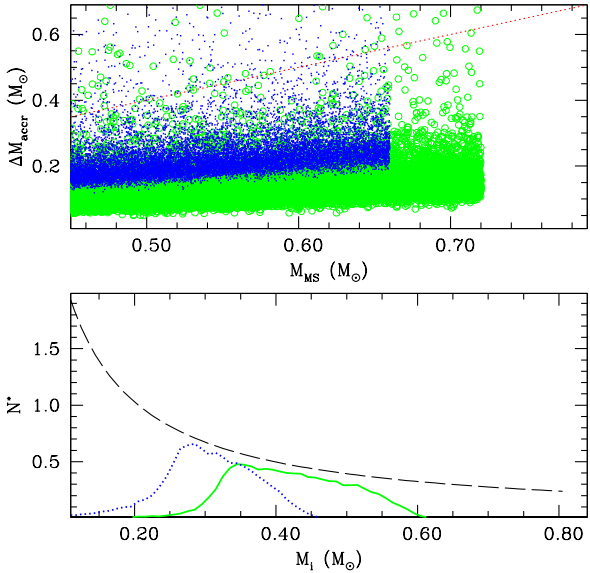}
\caption{As Fig.~\ref{sim1a} but for the case of variable ${\rm Y_{accr}}$, and ${\rm N(Y_{accr})\propto Y_{accr}^7}$ (see text for details).}
\label{sim3a}
\end{figure}

\begin{figure}
\centering
\includegraphics[scale=.4000]{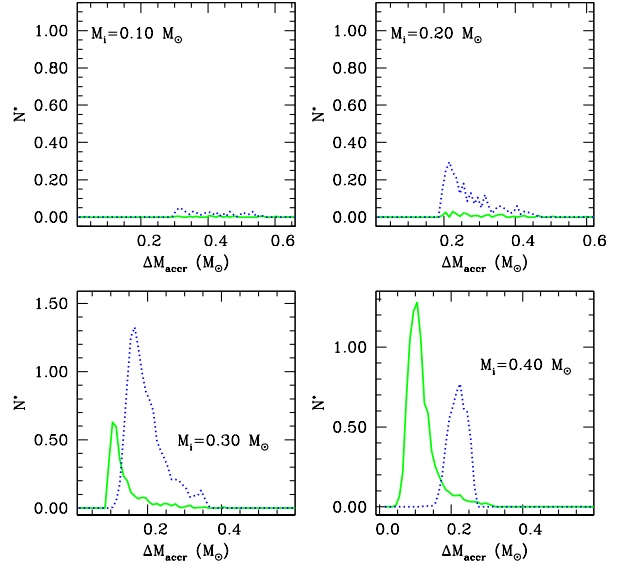}
\caption{As Fig.~\ref{sim1b}, but for the case ${\rm N(Y_{accr})\propto Y_{accr}^7}$ (see text for details).}
\label{sim3b}
\end{figure}

It is clear that in case of the assumed range of ${\rm Y_{accr}}$ values, the probability to have accreted matter with a given 
He mass fraction needs to be biased towards the largest ${\rm Y_{accr}}$ abundances. To this purpose we simulated a power law 
for the number of stars that have accreted matter with a given  ${\rm Y_{accr}}$, more specifically ${\rm N(Y) \propto Y^7}$. 
The results of this test are displayed in 
Figs.~\ref{sim3a} and ~\ref{sim3b}. About 3\% of potential bMS objects and less than 1\% of mMS stars spread over the 
whole sampled mass range cannot accrete enough matter 
to reach the appropriate final Y, number fractions that are probably still acceptable. Smaller exponents of the power law 
would be obviously ruled out.
The mean value of ${\rm Y_{accr}}$ is large, equal to 
0.57, with a 1$\sigma$ dispersion equal to 0.06. As for the mass budget, $\sim$44\% of the actual bMS mass 
and $\sim$24\% of the actual mMS mass must have been accreted.
Notice again the very different efficiency of accretion at fixed ${\rm M_i}$ between objects on the different sequences.

\subsection{The predicted CN and ONa anticorrelations}

\begin{figure}
\centering
\includegraphics[scale=.4000]{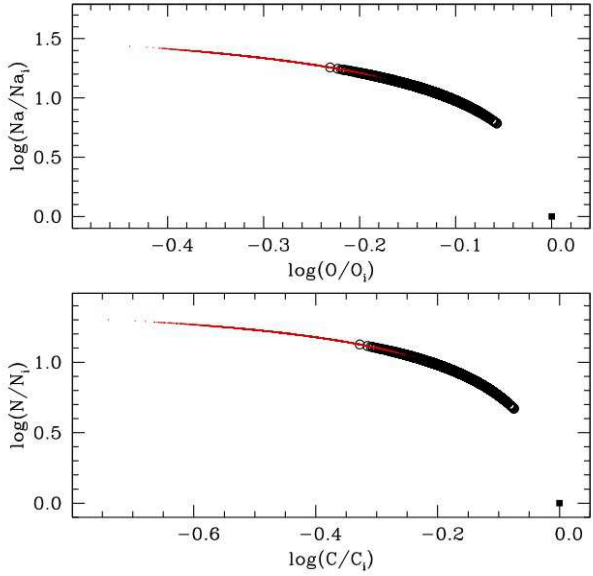}
\caption{C-N (upper panel) and O-Na (lower panel) anticorrelations (abundances are given as the logarithm of the ratio between 
the final and initial values) predicted by our simulations for the case with 
${\rm Y_{accr}=0.44}$ (see text for details). Red dots and black circles correspond to bMS and mMS objects, respectively. The filled squares 
displays the abundance ratios on the rMS.}
\label{anti_a}
\end{figure}

One of the difficulties with all scenarios to explain NGC2808 observations, is that 
whilst the He abundance seems to be somewhat {\sl quantized}, the heavy element anticorrelations 
(measured along the red giant branch), i.e. the O-Na anticorrelation 
\citep[see, e.g.,][]{carretta06} are not. 
Figure~\ref{anti_a} displays the pattern of C-N and O-Na abundances --in terms of ratios between the final and initial 
pre-accretion mass fractions-- obtained 
from our simulation with ${\rm Y_{accr}=0.44}$. The abundances in  
the accreted matter have been derived from the IMBS model displayed in Fig.~1 of \citet{demink} for Y$\sim$0.45 in the ejecta, assuming initial values from 
the $\alpha$-enhanced mixture of BaSTI models. 

\begin{figure}
\centering
\includegraphics[scale=.4000]{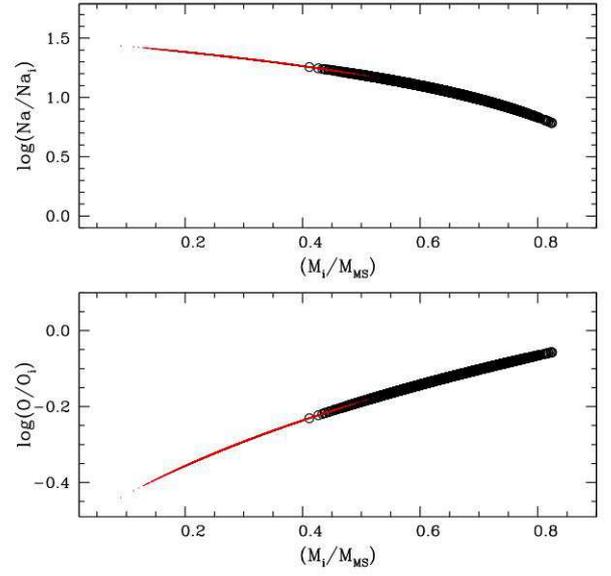}
\caption{As Fig.~\ref{anti_a}, but for the 
O and Na abundances as a function of the ratio between the initial and final MS mass (see text for details.}
\label{anti_b}
\end{figure}

Very interestingly, there is no quantization when moving from the mMS to the bMS, with a region of overlap 
between the abundances along the two sequence. This is an important result of this accretion scenario.
To explain this lack of discontinuity, 
it is instructive to examine Eq.~\ref{eq_b} and Fig.~\ref{anti_b}. 
Equation~\ref{eq_b} displays the relationship between  
the ratio of the final (${\rm X_f}$) to initial (${\rm X_i}$) mass fraction of a given element, the mass fraction in the 
accreted matter (${\rm X_{accr}}$), the initial and actual MS mass of a given star, e.g. 

\begin{equation}
{\rm 
\frac{X_f}{X_i}=\frac{M_i}{M_{MS}} + \frac{X_{accr}}{X_i} \ \left( 1- \frac{M_i}{M_{MS}} \right)
}
\label{eq_b}
\end{equation}

The ratio ${\rm M_i/M_{MS}}$ is fixed, for a given ${\rm Y_{accr}}$, by the actual 
Y abundances along the multiple MS, and displays a large range of values, reported 
in Fig.~\ref{anti_b} (for the case with ${\rm Y_{accr}=0.44}$) as a function of O and Na abundances. 
Given also that  ${\rm X_i}$, ${\rm Y_{accr}}$, and hence ${\rm X_{accr}}$, are common to both mMS and bMS, it is clear from 
Eq.~\ref{eq_b} that the overlapping range of ${\rm M_i/M_{MS}}$ ratios is the cause of the lack of discontinuity 
of the O and Na (also C and N) abundances when moving from the mMS to the bMS. 

A very important clarification is necessary here. When ${\rm Y_{accr}}$ is assumed constant during the accretion, 
the range of ${\rm M_i/M_{MS}}$ values is caused by the (small) intrinsic range of Y abundances of the mMS and bMS. 
If we assume Y to be just one single value along each of the two sequences, only two distinct values of ${\rm M_i/M_{MS}}$ 
would be possible, one for the mMS and one for the bMS, according to Eq.~\ref{eq}. In this case we would predict {\sl quantized} 
O-Na and C-N anticorrelations. 

The situation is different in the more realistic case of a range of ${\rm Y_{accr}}$ during the accretion, as discussed in the previous section. 
Even if Y is just one single value along each of the two sequences, Eq.~\ref{eq} shows clearly 
that a range of ${\rm Y_{accr}}$ abundances causes a range of ${\rm M_i/M_{MS}}$ ratios, thus producing a similar 
result to what is shown in Fig.~\ref{anti_b}.

There is however a discontinuity, mainly in the Na and N abundances, 
when moving from the rMS (that should display the initial values ${\rm X_i}$ for any element) 
and the mMS, contrary to the observations. 
The reason is related to the fact that the maximum value of the ratio (${\rm M_i/M_{MS}}$) is lower than one.
For the results displayed by Fig.~\ref{anti_b} this maximum is $\sim$0.8.
Substituting into Eq.~\ref{eq_b} gives 
${\rm (X_f/X_i) \sim 0.8 + (X_{accr}/X_i) * 0.2}$; given that ${\rm (X_{accr}/X_i)}\sim$20 for N and $\sim$30 for Na
\citep[from][model]{demink},  
we get minimum ${\rm (X_f/X_i)}$ ratios of $\sim$5 and $\sim$7 for N and Na, respectively. 
These minimum values are much closer to one for both C and O, because their abundances in the accreted matter are 
closer to the initial values, compared to N and Na.

All other simulations show also a maximum (${\rm M_i/M_{MS}}$) well below one, typically 0.7-0.8. This, coupled 
to the high values of ${\rm (X_{accr}/X_i)}$ for N and Na, always causes a discontinuity of a few 0.1~dex 
between the rMS and mMS abundances.  
 
Finally, we repeated our analysis by selecting only masses ${\rm M_{MS}}$ within 0.02~${\rm M_{\odot}}$ of the turn off, for both bMS and mMS. 
These objects correspond more closely to the stars currently on the red giant branch, for which spectroscopic measurements 
of, i.e., O and Na abundances are available. 
We found very similar results to those discussed above. They still display a range of O and Na abundances, an overlap between 
the abundance ranges spanned by bMS and mMS, and a discontinuity between rMS and mMS abundances. 
This was to be expected, given that for all simulations the range of (${\rm M_i/M_{MS}}$) ratios is approximately the same 
for all ${\rm M_{MS}}$ values, as can be inferred 
from Figs.~\ref{sim1a}, ~\ref{sim1ab}, ~\ref{sim1ac}, ~\ref{sim2a}, and ~\ref{sim3a}.

\section{Discussion and conclusions}

The goal of our analysis was to constrain the amount of accreted mass as a function 
of the actual MS mass, and He abundance of the accreted matter, which are required 
to reproduce the multiple MS observed in NGC~2808, within the framework of the early disc accretion scenario.
We have discussed separately the cases of constant and variable He abundance of the accreted matter.
Some interesting considerations arise from the analysis of our simulations:

\begin{itemize}

\item{In case of a constant ${\rm Y_{accr}}$, the minimum value to produce both mMS and bMS  
is $\sim0.44$. If ${\rm Y_{accr}}$ is below this limit, the bMS would appear truncated at luminosities 
where it is actually observed. The minimum constant ${\rm Y_{accr}}$ to produce just the mMS is $\sim$0.37.} 

\item{When considering a range 
of ${\rm Y_{accr}}$ values between 0.25 and 0.64, as for the IMBS model by \citet{demink}, our simulations 
show that a uniform distribution of the probability to accrete matter within this abundance range is ruled 
out. One needs instead a more {\sl tuned} distribution, strongly biased towards very large He mass fractions. 
As a general conclusion, very specific He abundance distributions must be accreted to produce the observed discrete sequences.}

\item{Regardless of the assumption on the He abundance distribution in the accreted matter, our simulations reveal that, at 
fixed initial mass of the accreting star, different efficiencies for the accretion are required, to produce stars placed 
onto the mMS or bMS. Within the early disc accretion scenario, this may be explained by 
differences in the orbital properties of the progenitors of mMS and bMS stars, and/or different spatial 
distribution of intracluster gas with varying He abundances.}

\item{Both O-Na and C-N anticorrelations appear naturally along the multiple MS, when considering the relationships between Y and 
CNONa abundances in the ejecta of the IMBS model. There is no discontinuity between the abundance ranges covered by mMS and 
bMS stars. This is a very important consequence of the early accretion scenario.}
There is however a sizeable (several 0.1~dex) discontinuity between rMS and mMS objects, which is related to the existence 
of an upper limit to the ${\rm M_i/M_{MS}}$ ratio, that is always well below one.

\end{itemize}

All these results have been obtained by assuming that the minimum mass of the accretors is 0.1${\rm M_{\odot}}$, e.g., 
approximately the minimum stellar mass. When employing lower values, in the brown dwarf regime, 
we found that for a constant ${\rm Y_{accr}}$, the minimum He mass fraction required 
to produce both mMS and bMS decreases slightly, by only a few percent, and approaches the limit 
${\rm Y_{accr}}\sim$0.40. Also, as easily gathered from 
Eq.~\ref{eq}, a decrease of ${\rm Y_{accr}}$, tends to shift to lower values the masses ${\rm M_i}$ of the accretors 
that have produced a star with a given actual mass ${\rm M_{MS}}$.
The general results for the case of a range of ${\rm Y_{accr}}$, and the appearance of the O-Na and C-N anticorrelations 
are all basically unchanged.

A second set of considerations need to address the question: do the ejecta of IMBS objects 
\citep[the major source of pollution  
for this scenario, according to][]{demink, bastian:13} have 
the correct He abundances to explain the observed mMS and bMS?
To address this issue we need to estimate the masses of both actual and accreted mass of mMS and bMS stars, the total 
amount of polluting matter available, and its He abundance distribution.
A source of uncertainty for these estimates 
is that the observed MFs of the multiple MS span a narrow 
mass range and have been determined just in one field. We do not have yet 
information about the radial distribution of the different stellar populations in NGC 2808, and   
moreover the present-day MF is the result of the cluster dynamical evolution, which has not been modelled. 

In the discussion that follows, we have estimated the mass budgets for the early accretion scenario assuming 
globally a standard Kroupa Initial MF \citep[IMF]{kroupa}. We also assumed that the observed population ratios of 
rMS, mMS and bMS stars are the same throughout the cluster, fixed at the end of the accretion process.
We focused on the mass range between $\sim$0.7 and $\sim$0.4${\rm M_{\odot}}$ for the mMS and bMS, covered by 
\citet{milone:12b} MF determinations, because 
we do not know yet whether the three separate sequences continue with the same number ratios to  
lower masses. The infrared data by \citet{milone:12a} display multiple sequences 
down to lower masses, but it is not clear the distinction between mMS and bMS, due to both photometric noise
and the lower sensitivity to He of the MS location of very low-mass stars.
It goes without saying that, if the three sequences were to be detected down to the H-burning limit, this 
would rule out the early accretion scenario.
We do not know the upper mass limit of the polluted stars 
(although in this scenario it cannot be much larger than ${\rm \sim1.5M_{\odot}}$, due to timescale arguments), 
and whether also masses evolved off the MS long ago 
were distributed exactly along the same three MSs we observe now. 

With these assumptions about IMF and mass ranges, 
and given that $\sim 24\%$ of MS stars belong to the mMS, and $\sim$14\% belong to the bMS \citep{milone:12b}, 
the former sequence contains, in the relevant mass range, $\sim$3\% of the total cluster mass, whereas the latter sequence contains 
$\sim$2\% of the total cluster mass.
The amount of available 
polluting matter is of the order of $\sim$13\% of the total cluster mass for IMBS \citep[assuming all massive stars 
above 10${\rm M_{\odot}}$ were born in interacting binary systems, as in][]{demink}.

We checked whether the 
model discussed in detail by \cite{demink} does eject enough matter with the right He abundances. 
In case of constant ${\rm Y_{accr}}$, our simulations with ${\rm Y_{accr}}$=0.44 --the minimum 
He mass fraction required to produce both sequences-- dictate that $\sim$2.5\% 
(shared in almost equal amounts between the two sequences) of the total cluster mass 
must be accreted with this abundance. 
Figure~1 in \citet{demink} shows that $\sim 15$\% of the ejected gas (the last $\sim 1.5 {\rm M_{\odot}}$ 
of gas ejected by a given IMBS) display a He mass fraction larger than 
$\sim$0.45; this translates into $\sim$2\% of the total cluster mass. This number is close to what required 
by our simulation, especially when taking into account that we are considering matter ejected  
with ${\rm Y_{accr}}$=0.45 as a lower limit. 
Given that higher ${\rm Y_{accr}}$ requires less accretion, the model by 
\citet{demink} appears to be able to produce enough polluting matter with the appropriate He abundances to reproduce the observed 
portions of the multiple MS. 
Of course the issue here is that a mechanism is needed to explain why all gas previously ejected with a lower 
Y --$\sim$85\% of the total ejected mass-- cannot be accreted.
It is implicit the important constraint that the ejecta of these IMBS systems cannot be fully mixed before being accreted, 
otherwise the He abundances (the averaged mass fraction would be $\sim$0.30) are not 
high enough to explain the observed multiple MS. 

When we consider our simulation with variable ${\rm Y_{accr}}$ that is able to reproduce the multiple MS 
(${\rm N(Y) \propto Y^7}$), we obtain that, as expected,   
negligible fractions of the accreted matter have ${\rm Y_{accr}}<$0.45. A fraction $\sim$1.4\% of the total cluster mass 
needs to be accreted with Y above 0.45, a figure still consistent with the ejected 
mass budget of IMBS models. Moreover, 
this simulation requires that the accreted matter is strongly biased towards high ${\rm Y_{accr}}$ (0.64 is the upper limit 
we considered) and therefore we evaluated
also the amount of accreted mass with ${\rm Y_{accr}}>$0.60. This turned out to be $\sim$0.5\% of the total 
cluster mass. The IMBS model predicts a value equal to $\sim$0.6\%, again consistent with the requirements.  

All these results are basically unchanged if we consider minimum masses for the accretors in the brown dwarf regime.

In conclusion:

\begin{itemize}

\item{
The early accretion scenario is quantitatively compatible with the observed He abundances of the multiple sequences 
of NGC2808, at least within the mass ranges where the multiple MS is currently well defined 
(lower mass limit of $\sim$0.4${\rm M_{\odot}}$ for mMS and bMS.}

\item{According to the models by \citet{demink}, IMBS objects should eject enough mass with the right Y abundance to explain the observed 
multiple MS. But a mechanism is required, whereby the first (in terms of time of the ejection) 
$\sim$85\% of the total mass released by an IMBS is not accreted, because only the last 15\% display the right range of Y values.  
This is the matter ejected at the end of the first phase of mass transfer 
\citep[after central H exhaustion of the primary, see][]{demink}, and during the second phase of mass transfer, which coincides with the shell He-burning stage
of the primary (by now a Wolf-Rayet star). 
It is important that the matter ejected by the IMBS polluters is not fully mixed before being accreted otherwise, according to the models, the average Y abundance 
of the polluting gas is only $\sim$0.30, too low to produce the observed multiple MS.}

\item{So far there exists just one detailed IMBS simulation. It is very important to explore the parameter space not only regarding masses of the
binary components and their separation, but also in terms of uncertainties of the binary model calculations, 
such as mass transfer efficiency and mixing processes, to predict accurately the possible range of mass and chemical composition of their ejecta.}

\item{Because of the effect of dynamical evolution, a global observational analysis of the cluster is necessary, to assess whether 
over the whole cluster the multiple sequences continue to be well separated, and the population ratios conserved,   
down to the H-burning limit. 
If this is the case, the early accretion scenario would encounter difficulties, given that 
in this model the stars on bMS and mMS must be originated 
by accretion on objects originally on the rMS.
The reason stems from the fact that the maximum 
amount of accreted mass for a given ${\rm M_{MS}}$ is fixed by the composition of the ejecta (Eq.~\ref{eq}), whereas the minimum allowed value of ${\rm Y_{accr}}$ is determined by the lower limit on ${\rm M_i}$.
Even for the extreme case ${\rm Y_{accr}}$=0.40 (minimum ${\rm M_i}$ equal to zero) stars along the mMS 
with mass above $\sim$0.2${\rm M_{\odot}}$ are originated by accretors with ${\rm M_i}$ above $\sim$0.1${\rm M_{\odot}}$, in the stellar regime. 

At the same time, full numerical simulations of the dynamics of the cluster stars, the distribution and chemical composition 
of the polluting gas, and the efficiency of disc accretion are necessary 
to assess whether the early disc accretion scenario can satisfy all the constraints discussed here.
}

\item{The multiple populations hosted by NGC2808 reach very high initial 
He abundances compared to other GGCs. These high abundances, their {\sl quantization}, and the 
seemingly {\sl continuous} distribution of O-Na anticorrelations, are a challenge to 
all scenarios proposed to explain the multipopulation phenomenon. 
We have studied here whether/under which conditions the early disc accretion scenario 
can satisfy these observational constraints. 
It is paramount to extend this type of investigation to other GGCs where the initial 
He abundances of the various populations can be determined,  
to provide further constraints/tests for this scenario.}

\end{itemize}


\begin{acknowledgements}
We thank our anonymous referee for a very prompt report and comments that 
have improved the presentation of our results.
SC is grateful for financial support
from PRIN-INAF 2011 "Multiple Populations in Globular Clusters: their
role in the Galaxy assembly" (PI: E. Carretta), and from PRIN MIUR 2010-2011,
project \lq{The Chemical and Dynamical Evolution of the Milky Way and Local Group Galaxies}\rq, prot. 2010LY5N2T (PI: F. Matteucci). 
\end{acknowledgements}

\bibliographystyle{aa}
\bibliography{N2808accr}

\end{document}